\documentclass[pdf,prd,aps,preprintnumbers,nofootinbib]{revtex4}

\usepackage[normalem]{ulem}

\usepackage{epsfig}
\usepackage{amsmath}
\usepackage{hyperref}
\usepackage{xspace}
\usepackage{color}

\newcommand{\mrm}[1]{\mathrm{#1}}
\newcommand{\ttt}[1]{\texttt{#1}}
\newcommand{\py}{\ttt{PYTHIA}}
\newcommand{\hw}{\ttt{HERWIG}}
\newcommand{\sh}{\ttt{SHERPA}}
\newcommand{\pho}{\ttt{PHOJET}}
\newcommand{\sib}{\ttt{SIBYLL}}
\newcommand{\qgs}{\ttt{QGSJET}}
\newcommand{\epos}{\ttt{EPOS}}
\renewcommand{\sh}{\ttt{SHERPA}}

\newcommand{\figRef}[1]{fig.~\ref{#1}}

\begin{document}

\renewcommand{\thefigure}{\arabic{figure}}

\title{Soft-QCD and UE spectra in pp collisions at very high CM
  energies\\(a Snowmass white paper)} 

\author{Peter Skands}
\affiliation{Theory Division, CERN, CH-1211 Geneva 23, Switzerland}

\date{August 13, 2013}

\begin{abstract}
We make some educated guesses for the extrapolations of typical 
soft-inclusive (minimum-bias, pileup, underlying-event) observables 
to proton-proton collisions at center-of-mass energies in the range 13 --
100 TeV. The numbers should be interpreted with (at least) a $\pm 10
\%$ uncertainty.
\end{abstract}
\maketitle

\section{Soft Physics Models and Energy Scaling}
Soft physics models can essentially be divided into two broad
categories. The first starts from perturbative QCD (partons, matrix
elements, jets) and uses a factorized
perturbative expansion for the hardest parton-parton interaction, 
combined with detailed models of hadronization and (soft and hard) 
multiparton interactions (MPI). This is the 
approach taken by general-purpose event generators,
like \hw~\cite{Corcella:2000bw,Bahr:2008pv},
\py~\cite{Sjostrand:2006za,Sjostrand:2007gs}, and
\sh~\cite{Gleisberg:2008ta}. Since they agree with perturbative QCD
(pQCD) at high $p_\perp$, they are used extensively by the
collider-physics community, 
see \cite{Buckley:2011ms,Skands:2011pf} for reviews. 
The price is a typically reduced predictivity for very soft
physics. Collisions involving nuclei with $A\ge 2$ are generally not
addressed at all by these generators, though extensions
exist~\cite{Armesto:2009fj,Gyulassy:1994ew}.  

At the other end of the spectrum are tools starting from Regge
theory (optical theorem, cut and uncut pomerons), like
\qgs~\cite{Ostapchenko:2004ss} and 
\sib~\cite{Ahn:2009wx}. These are typically used
e.g.\ for cosmic-ray air showers and for heavy-ion collisions. 
The main focus is here on the soft physics, though perturbative
contributions can be added in, e.g.\ by the introduction of a ``hard
pomeron''. Inbetween are tools like \pho~\cite{Bopp:1998rc},
\ttt{DPMJET}~\cite{Bopp:2005cr}, and \epos~\cite{Werner:2010aa}, which contain
elements of both 
languages (with \epos\ adding a further component:
hydrodynamics~\cite{Werner:2007bf}).  
Note, however, that all of these models rely on string models of
hadronization and hence have some overlap with \py\ on that aspect of
the event modeling.

The educated guesses in this summary are based mainly on the energy scaling
exhibited by the Perugia 2012 set of tunes~\cite{Skands:2010ak} 
of the $p_\perp$-ordered MPI model~\cite{Sjostrand:2004pf,Sjostrand:2004ef} 
in the \py~6 generator,  
which have been validated to give an acceptable description of the scaling of
a wide range of min-bias and underlying-event  
observables from lower collider energies to the
LHC~\cite{Skands:2010ak,Karneyeu:2013aha} (see e.g., 
\url{mcplots.cern.ch}~\cite{Buckley:2010ar,Karneyeu:2013aha}). The set 
also includes several theory uncertainty variations (e.g.,
of renormalization scales, PDFs, IR cutoffs, and color reconnections).    
For minimum-bias
and pile-up type observables, we also draw on comparisons to
\epos, \sib, and \qgs, taken from the study in~\cite{d'Enterria:2011kw}.

\begin{figure}[t]
\centering
\begin{tabular}{rc}
\rotatebox{90}{\hspace*{1.5cm}\small$p_{\perp 0}~[\mrm{GeV}]$} &
\includegraphics*[scale=0.65]{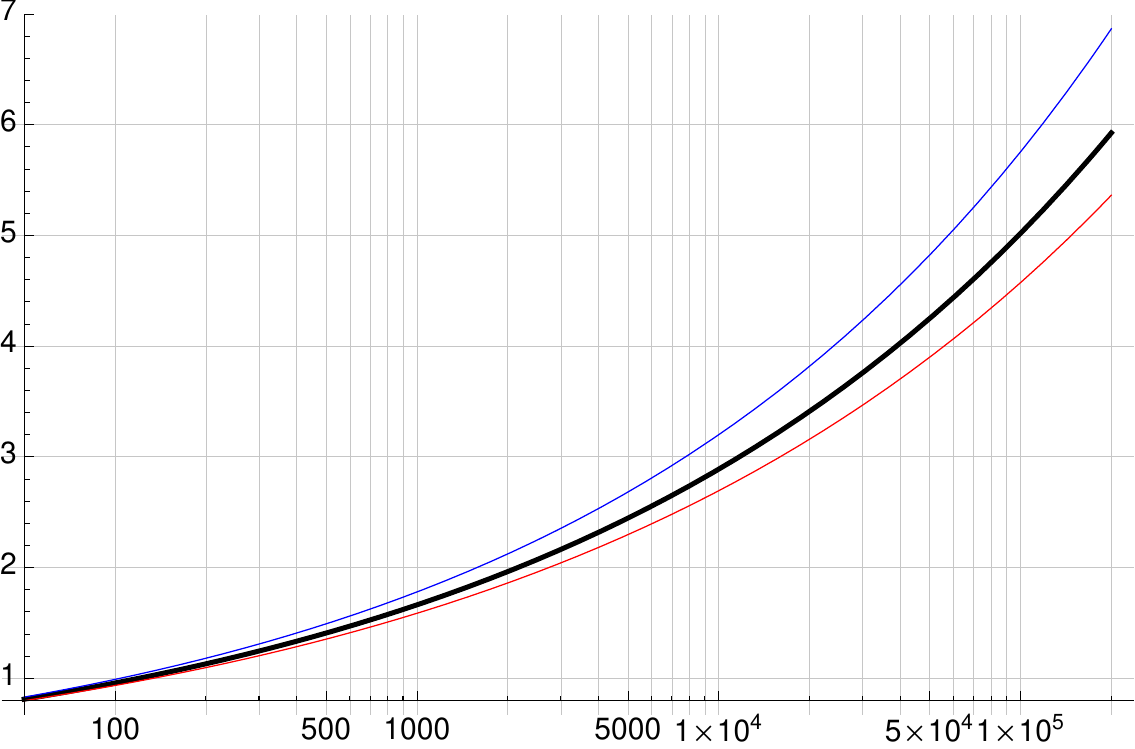}\\
&\small $E_\mrm{CM}~[\mrm{GeV}]$
\end{tabular}
 \hspace*{1cm} 
\begin{tabular}{c}\includegraphics*[scale=0.55]{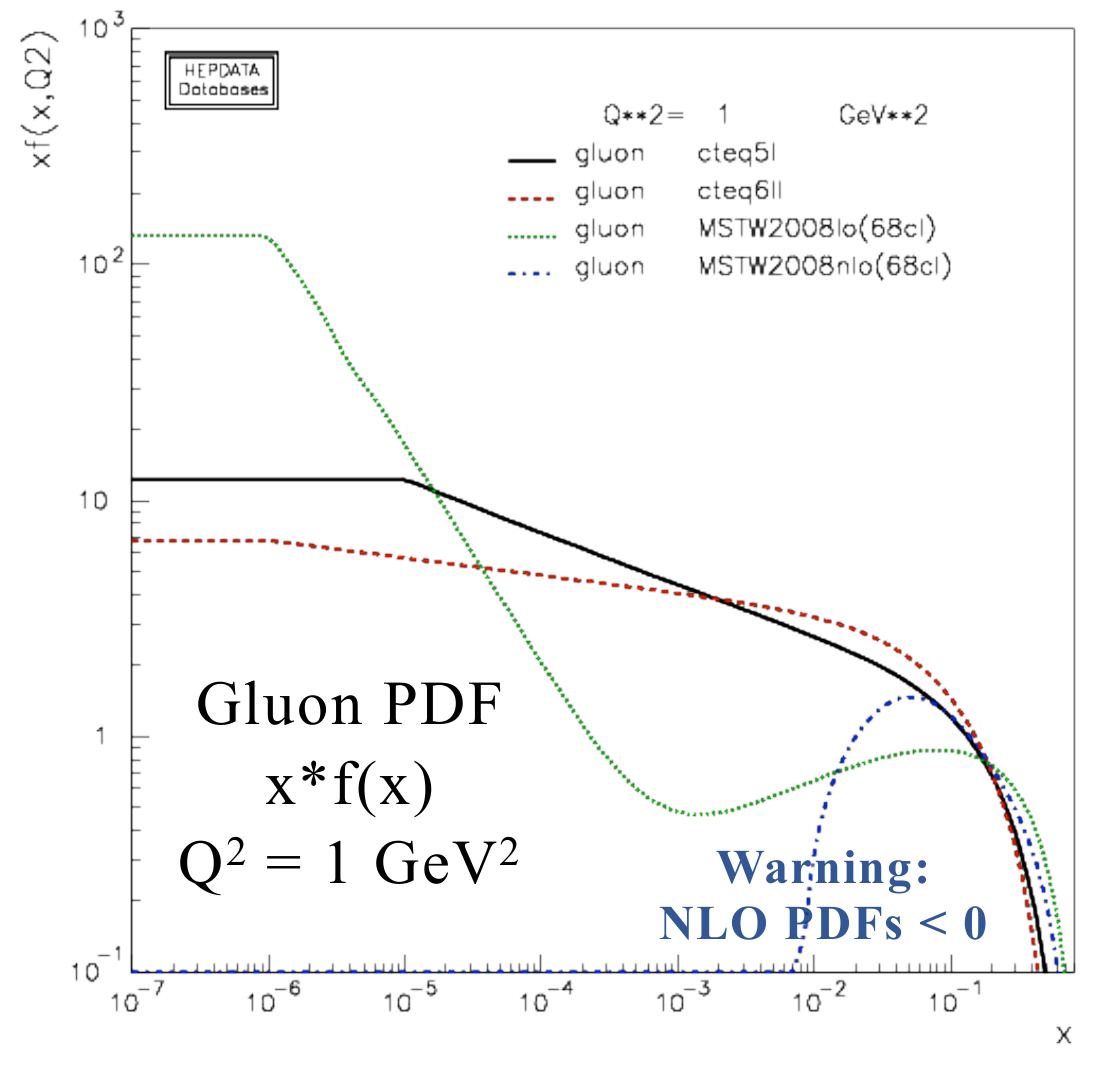} 
\end{tabular}
\caption{{\sl Left:} energy scaling of
  the $p_{\perp 0}$ soft-MPI regularization scale in the Perugia tunes
  (central value and range).
{\sl Right:} behavior of some typical PDF sets at very low
  $Q^2 = 1\,\mrm{GeV}^2$, plot from
  \ttt{HEPDATA}~\cite{Buckley:2010jn}. 
 \label{fig:PDFs}}
\end{figure}

In MPI-based models, one should be aware that 
the amount of soft MPI is sensitive to the 
PDFs at low $x$ and $Q^2$, a region which is not especially well
controlled. Physically, color screening and/or saturation effects
should be important. In practice, one introduces 
an $E_\mrm{CM}$-dependent regularization scale, $p_{\perp 0}$, illustrated for
the Perugia models in the left-hand pane of \figRef{fig:PDFs}. There
is then still 
a dependence on the low-$x$ behavior of the
PDF around that scale, illustrated 
in the right-hand pane. 
Note the freezing of the PDFs at very low $x$ 
(only marginally relevant for $E_\mrm{CM}\le 100\,\mrm{TeV}$). 
Note also that NLO PDFs should not be used for MPI models, since they
are not probability densities (e.g., they can become negative, illustrated
here by the MSTW2008 NLO set~\cite{Martin:2009iq}). The Perugia 2012
tunes are based on the CTEQ6L1 LO PDF set~\cite{Pumplin:2002vw},
but include MSTW2008 LO~\cite{Martin:2009iq} and MRST
LO**~\cite{Sherstnev:2008dm} variations.  

The issue of final-state color reconnections (CR) 
is another important aspect which is less than optimally
understood. Physically, this may reflect 
a generalization of color coherence, and/or dense-packing effects
(parton-, string-, or hadron-rescattering). In practice, the
Perugia tunes employ a nonperturbative CR model which is
based on the string area law~\cite{Sandhoff:2005jh,Skands:2007zg}. 
There are indications that
higher CM energies may require a smaller effective CR
strength~\cite{Schulz:2011qy}. 
If so, multiplicities could effectively increase a bit faster than we
assume here, a possibility we take into account when we evaluate the
extrapolation uncertainties. 

\section{Extrapolations to Very High Energies}

\begin{figure}[t]
\centering
\begin{tabular}{rc}
\rotatebox{90}{\hspace*{1.3cm}\small Cross Section, $\sigma$} &
\includegraphics*[scale=0.5]{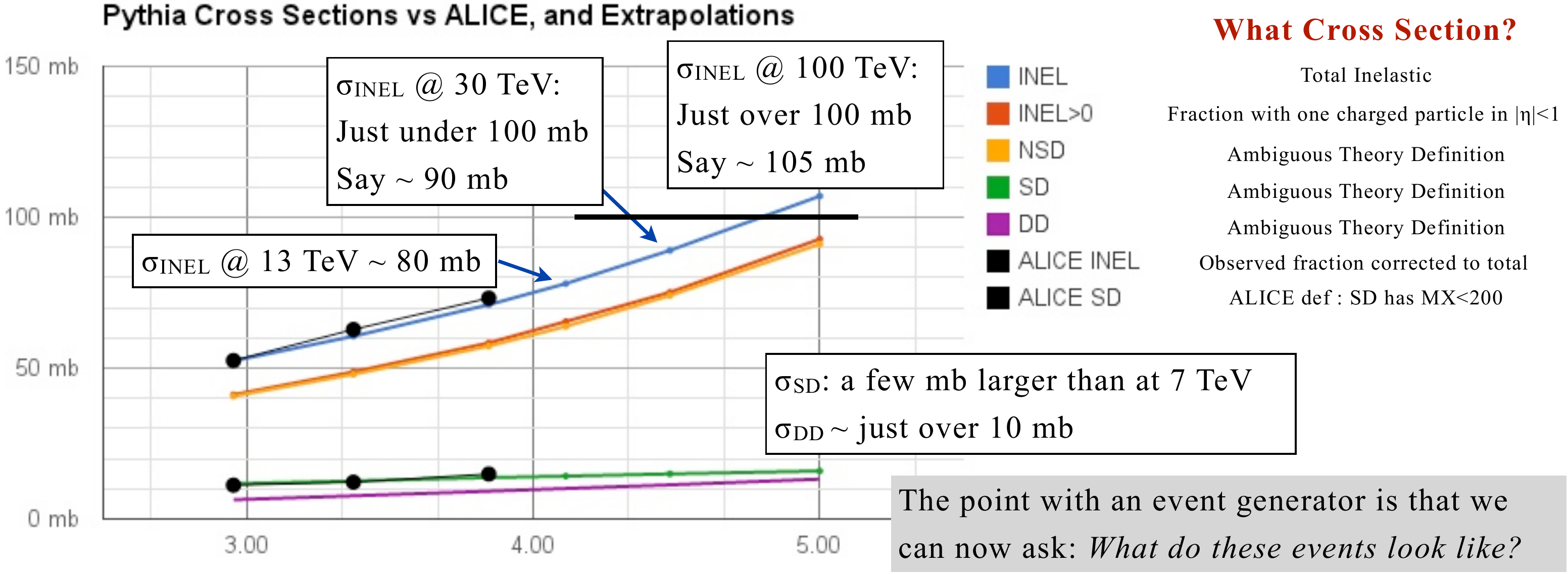}\\
& $\log_{10}(E_\mrm{CM}/\mrm{GeV})$\hspace*{5cm} \ 
\end{tabular}
\caption{Assumed scaling of various inelastic components of the total
  cross section in \py~6.\label{fig:py6sigmas}}
\end{figure}

For the total cross section, we take a simple Donnachie-Landshoff fit with
$\epsilon\sim 0.08$~\cite{Donnachie:1992ny}, 
which is also the basis of the scaling ans\"atze made in
\py~\cite{Schuler:1993wr}. As can be seen in \figRef{fig:py6sigmas},
ALICE measurements of the inelastic and single-diffractive cross
sections~\cite{Abelev:2012sea} exhibit no significant deviations from
this ansatz over the measured range. 
Note that the various diffractive components cannot be unambiguously
defined without a specific observable definition (supplied by ALICE as
a cut on the mass of the diffractive system, $M_X<200\,\mrm{MeV}$). 
The extrapolations yield an inelastic cross section growing from
$sim 80$ mb at 13 TeV to $\sim$ 90 mb at 30 
TeV and $\sim$ 105 mb at 100 TeV, while the elastic cross section (not
shown) increases from about 22 mb to 25 mb and 32 mb in the same range. The
diffractive components increase by only a few mb relative to their 7-TeV
values. 
We can now take a closer look at what these collisions look like. How
many tracks, and how much energy deposition are they associated with? 

\begin{figure}[t]
\centering
\begin{tabular}{c}
 \includegraphics*[scale=0.38]{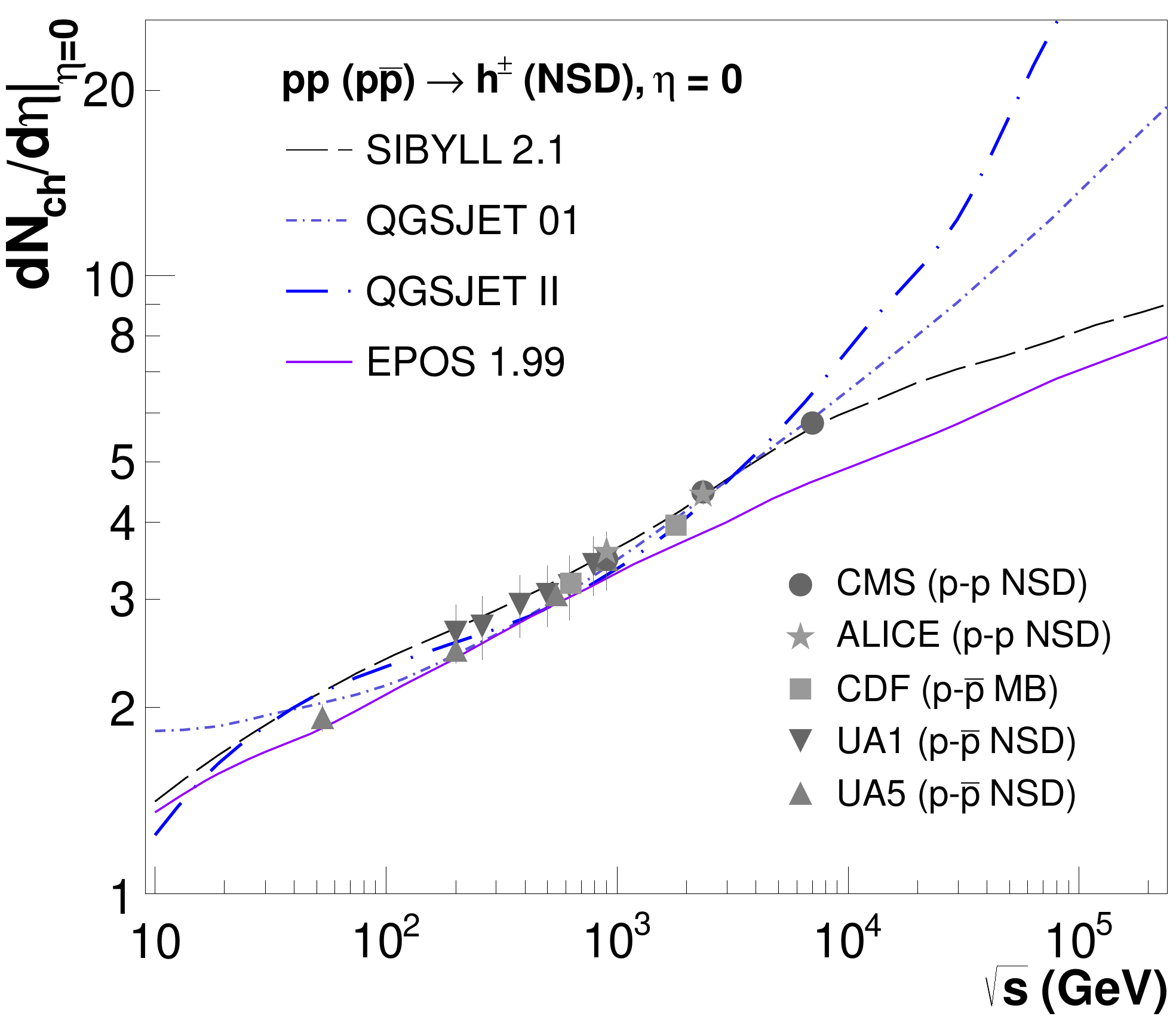}
\end{tabular}
\hspace*{1cm}
{\small\begin{tabular}{c}
\sl Central Charged-Track Multiplicity\\
\sl\bf Relative Increase:\\ 
Open Squares: 0.9 -- 2.36 TeV\\
Solid Squares: 0.9 -- 7 TeV\\
Grey Bands: ALICE\\
\includegraphics*[scale=0.4]{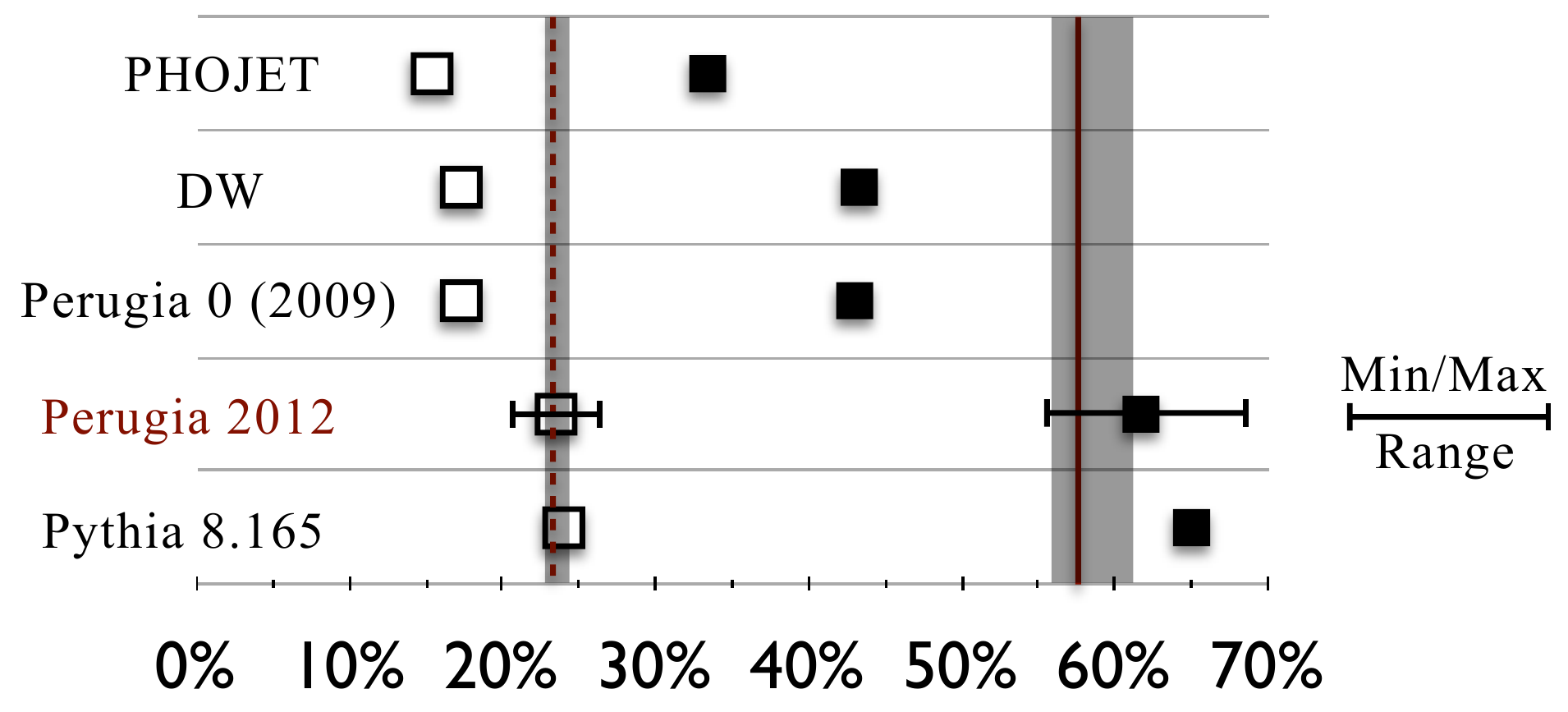}
\end{tabular}}
\caption{{\sl Left:} scaling of the central charged multiplicity for
  \sib, \qgs, and \epos, compared with collider data for NSD
  events, from~\cite{d'Enterria:2011kw}. {\sl Right:} updated version 
of a plot in~\cite{Aamodt:2010pp} including present-day \py~6 and 8
tunes. 
\label{fig:dndeta}
}
\end{figure}

\begin{figure}[tp]
\centering
\begin{tabular}{rl}
&\hspace*{0.5cm}\includegraphics*[scale=0.53]{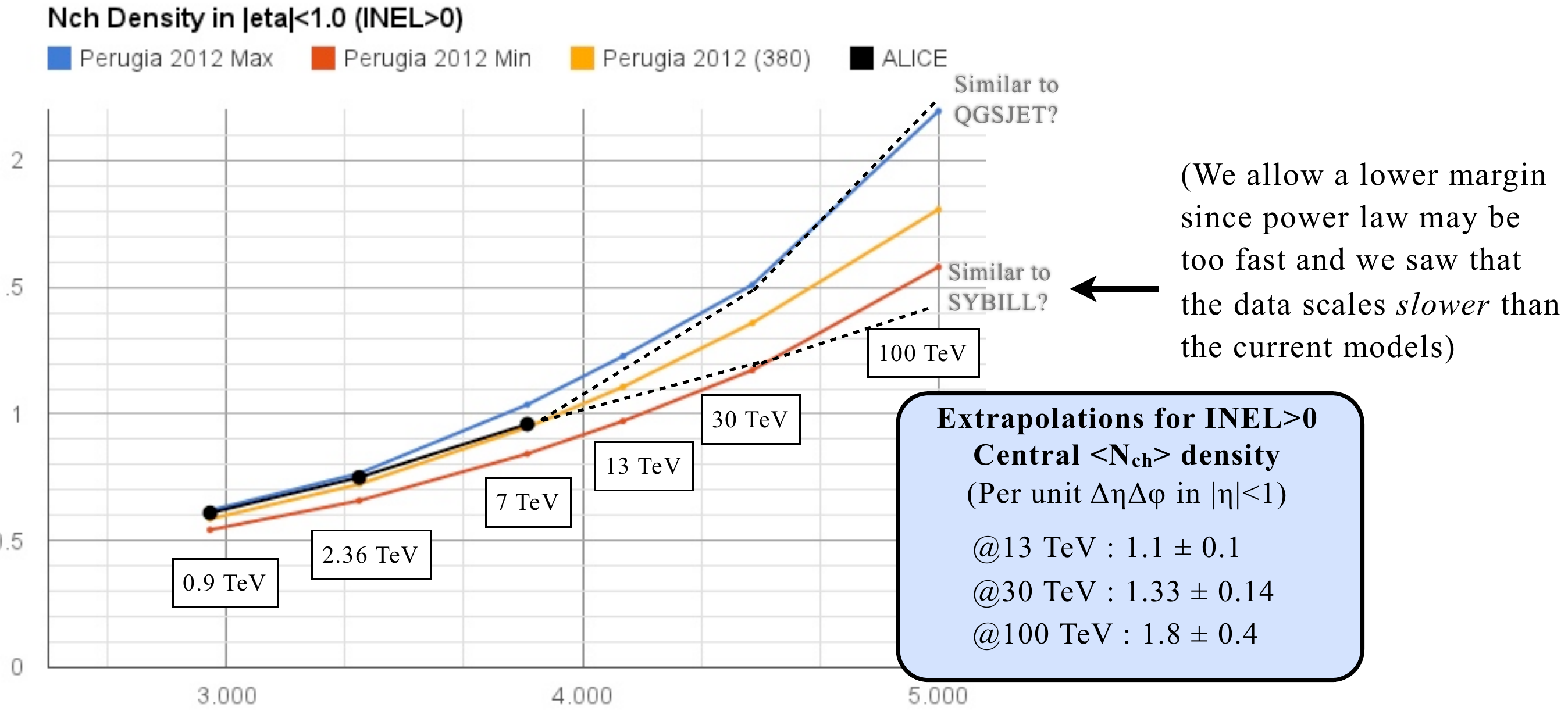}\\
& \small\hspace*{4.6cm}$\log_{10}(E_\mrm{CM}/\mrm{GeV})$\\[3mm]
&\includegraphics*[scale=0.5]{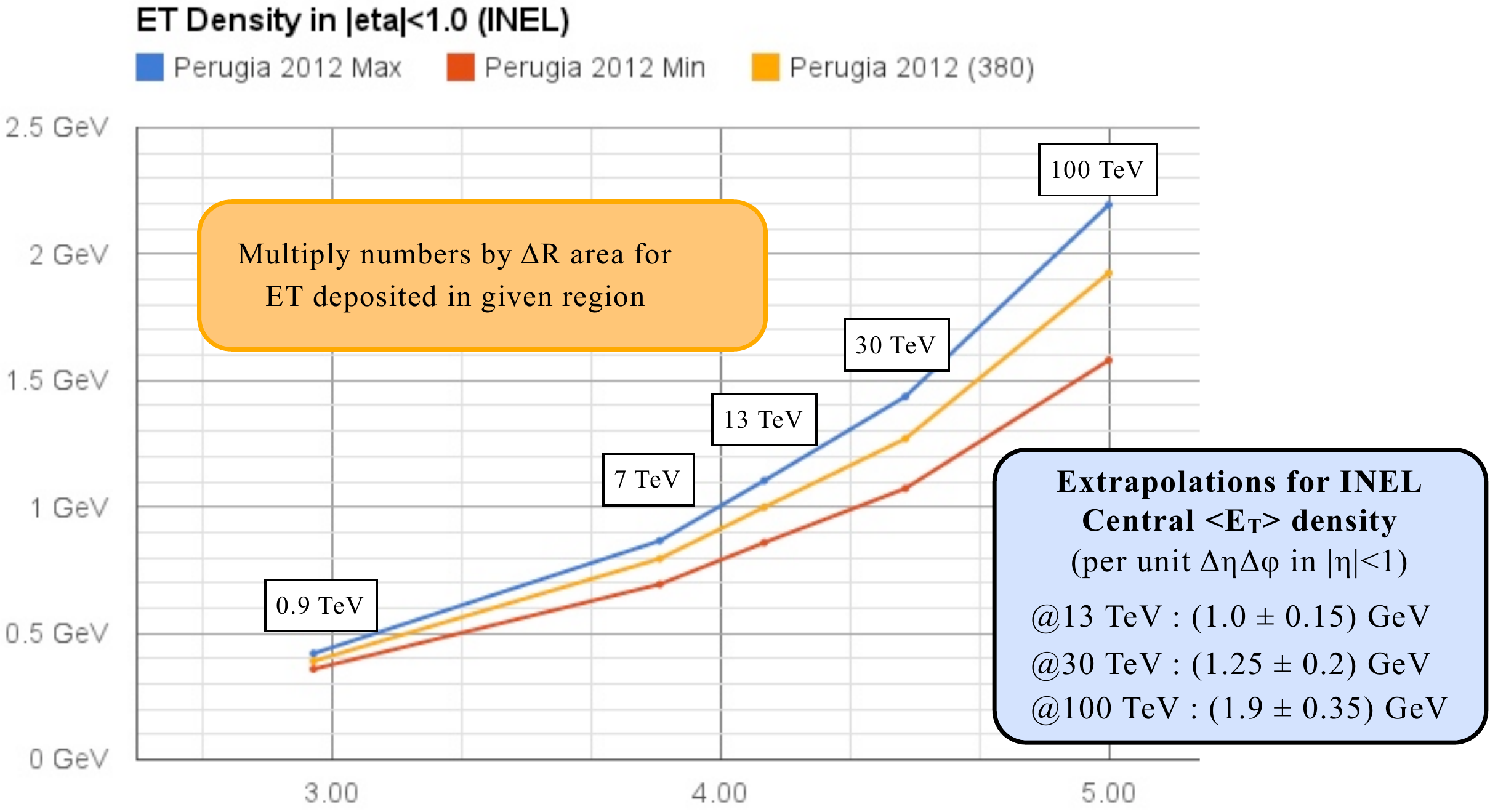}\\
& \small\hspace*{4.6cm}$\log_{10}(E_\mrm{CM}/\mrm{GeV})$\\[3mm]
&\hspace*{0.05cm}\includegraphics*[scale=0.58]{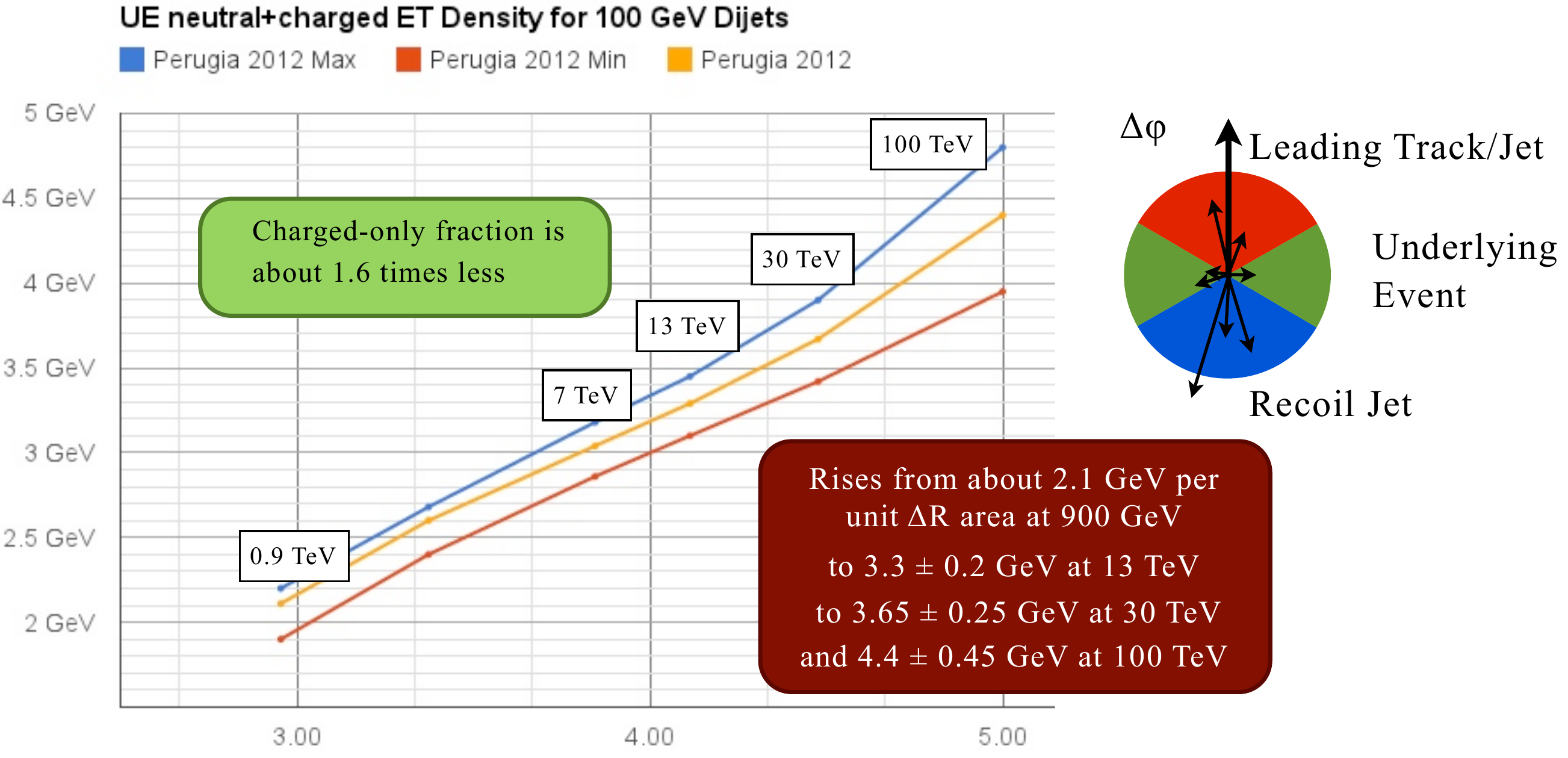}\\
& \small\hspace*{4.6cm}$\log_{10}(E_\mrm{CM}/\mrm{GeV})$
\end{tabular}
\caption{Extrapolations of the central ($|\eta|<1$) charged-track density for
  INEL$>$0 events ({\sl
    Top}), central ($|\eta|<1$) $E_T$ density for INEL events ({\sl
    Middle}), and central  ($|\eta|<2.5$) 
  UE $E_T$ density for 100-GeV dijet events ({\sl Bottom}). 
\label{fig:extrapolations}}
\end{figure}

Extrapolations of central charged-track densities in so-called
non-single-diffractive events in pomeron-based models are shown in 
the left-hand pane of \figRef{fig:dndeta}, from
\cite{d'Enterria:2011kw}.  
We note that the version of \epos\ used in \cite{d'Enterria:2011kw}
predicts a much too slow rise with CM energy, while \qgs~\ttt{II} errs
severely in the opposite direction, thus we exclude them from our
estimates. 
The \pho\ and \py\ generators are represented
on the right-hand pane of
\figRef{fig:dndeta}, which contains an update of a highly sensitive
plot made by the ALICE
collaboration~\cite{Aamodt:2010ft,Aamodt:2010pp}. It shows the 
relative increase in central charged-track multiplicity 
between 900 GeV and the 2360 and 7000 GeV CM
energies at the LHC, for events with at least one charged track inside
$|\eta|<1$ (INEL$>$0). Both \pho\ and the
Tevatron tunes of \py~6 (DW and Perugia 0) exhibit too slow
increases with energy, and hence 
are not included in our extrapolations. The Perugia 2012 and \py~8
(tune 4C~\cite{Corke:2010yf}) models, 
however, manage to reproduce the scaling observed by ALICE fairly
well. They can therefore be used as a reasonable first guess for 
further extrapolations, illustrated in the top pane of
\figRef{fig:extrapolations}. 
Combining the Perugia uncertainty variations
with the \sib and \qgs~\ttt{01} scaling trends yields an estimated 
central charged-track density per unit $\Delta\eta\Delta\phi$ of 
$1.1\pm0.1$ at 13 TeV, $1.33\pm 0.14$ at 30 TeV, and $1.8\pm0.4$ at
100 TeV, for inelastic 
events with at least one track inside $|\eta|<1$ (corresponding to the
red cross-section curve in \figRef{fig:py6sigmas}).  

Note that, when imposing $p_\perp$ cuts on the tracks, 
one should be aware that indications
from the LHC so far are that the $p_\perp$ spectra produced by
\py\ are slightly too hard~\cite{Karneyeu:2013aha}, with a deficit of
about 20\% for  
$p_\perp$ values below $\sim$ 200 MeV, and a similar excess above
$\sim$ 4 GeV. (This applies to inclusive charged tracks. Uncertainties 
are substantially larger for identified particles.)
Going from $|\eta|<1$ to $|\eta|\le 3$, say, 
does not change these predictions considerably. There is the trivial
seagull-shaped pseudorapidity distribution (roughly a 10\% effect),
but no other major differences in estimated track densities or spectra. 

An important quantity for jet energy scale calibrations is the amount
of transverse energy deposited in the detector, per unit
$\Delta R^2=\Delta\eta\times\Delta\phi$, per inelastic collision
(corresponding to the blue cross-section curve in
\figRef{fig:py6sigmas}). 
In the central region
of the detector, the Perugia models are
in good agreement with ATLAS measurements at 7
TeV~\cite{Aad:2012mfa,Karneyeu:2013aha}, while the activity in the 
forward region is
underestimated~\cite{Chatrchyan:2011wm,Aad:2012mfa,Aspell:2012ux,Karneyeu:2013aha}.  
Extrapolations lead to an estimated
$1.0 \pm 0.15\,\mrm{GeV}$ of transverse energy deposited per unit
$\Delta R^2$ in the central region of the detector at 30 TeV, growing
to 
$1.25\pm0.2\,\mrm{GeV}$ at 30 TeV, and 
$1.9\pm0.35\,\mrm{GeV}$ at 100 TeV, shown in the middle pane of
\figRef{fig:extrapolations}. 
We emphasize that similar
extrapolations in the forward region would likely result in
underestimates by up to a factor 1.5, at least if done with current
\py\ models. 

The last quantity we consider is the activity in the underlying
event (UE). The most important UE observable is the summed
$p_\perp$ density in the so-called ``TRANSVERSE'' region, defined as
the wedge $60-120^\circ$ away in azimuth from a hard trigger jet. For
$p_{\perp}^\mrm{jet}$ values above 5 -- 10 GeV, this
distribution is effectively flat, i.e., to first approximation 
it is independent of the jet $p_\perp$. It does, however,
depend significantly on the CM energy of the pp
collision, a feature which places strong constraints on the scaling of
the $p_{\perp0}$ scale of MPI models, cf.~\figRef{fig:PDFs}. 
Given the good agreement between the Perugia 2012 models and Tevatron
and LHC UE measurements~\cite{Karneyeu:2013aha}, we estimate the  
$E_T$ (neutral+charged) density in the TRANSVERSE region (inside
$|\eta|<2.5$), for a reference case of 100-GeV dijets in the bottom
pane of \figRef{fig:extrapolations}: starting 
from an average of about $2.1\,\mrm{GeV}$ per
$\Delta R^2$ at 900 GeV, the density 
rises to $3.3 \pm 0.2\,\mrm{GeV}$ at 13 TeV, $3.65\pm 0.25\,\mrm{GeV}$ 
at 30 TeV, and $4.4\pm 0.45\,\mrm{GeV}$ at 100 TeV. Note that the
charged-only fraction of this would be about a factor 1.6 less. 

\subsection*{Acknowledgments}
We are grateful to the volunteers
participating in the LHC@home 2.0 project
``Test4Theory''~\cite{LombranaGonzalez:2012gd} for
the CPU time they contribute to the \url{mcplots.cern.ch} web
site~\cite{Karneyeu:2013aha}, used in this work.  Also many thanks to
D.~d'Enterria for providing the plot shown in the left-hand pane of 
\figRef{fig:dndeta}.

\bibliographystyle{utphys}
\bibliography{minbias.bib}

\providecommand{\href}[2]{#2}\begingroup\raggedright\begin{thebibliography}{10}

\bibitem{Corcella:2000bw}
G.~Corcella, I.~Knowles, G.~Marchesini, S.~Moretti, K.~Odagiri, {\em et al.},
  ``{HERWIG 6: An Event generator for hadron emission reactions with
  interfering gluons (including supersymmetric processes)},'' {\em JHEP} {\bf
  0101} (2001) 010,
\href{http://www.arXiv.org/abs/hep-ph/0011363}{{\tt hep-ph/0011363}}.

\bibitem{Bahr:2008pv}
M.~B{\"a}hr, S.~Gieseke, M.~Gigg, D.~Grellscheid, K.~Hamilton, {\em et al.},
  ``{Herwig++ Physics and Manual},'' {\em Eur.Phys.J.} {\bf C58} (2008)
  639--707,
\href{http://www.arXiv.org/abs/0803.0883}{{\tt 0803.0883}}.

\bibitem{Sjostrand:2006za}
T.~Sj{\"o}strand, S.~Mrenna, and P.~Z. Skands, ``{PYTHIA 6.4 Physics and
  Manual},'' {\em JHEP} {\bf 0605} (2006) 026,
\href{http://www.arXiv.org/abs/hep-ph/0603175}{{\tt hep-ph/0603175}}.

\bibitem{Sjostrand:2007gs}
T.~Sj{\"o}strand, S.~Mrenna, and P.~Z. Skands, ``{A Brief Introduction to
  PYTHIA 8.1},'' {\em Comput.Phys.Commun.} {\bf 178} (2008) 852--867,
\href{http://www.arXiv.org/abs/0710.3820}{{\tt 0710.3820}}.

\bibitem{Gleisberg:2008ta}
T.~Gleisberg, S.~H{\"o}che, F.~Krauss, M.~Sch{\"o}nherr, S.~Schumann, {\em et
  al.}, ``{Event generation with SHERPA 1.1},'' {\em JHEP} {\bf 0902} (2009)
  007,
\href{http://www.arXiv.org/abs/0811.4622}{{\tt 0811.4622}}.

\bibitem{Buckley:2011ms}
A.~Buckley, J.~Butterworth, S.~Gieseke, D.~Grellscheid, S.~H{\"o}che, {\em et
  al.}, ``{General-purpose event generators for LHC physics},'' {\em
  Phys.Rept.} {\bf 504} (2011) 145--233,
\href{http://www.arXiv.org/abs/1101.2599}{{\tt 1101.2599}}.

\bibitem{Skands:2011pf}
P.~Z. Skands, ``{QCD for Collider Physics},''
\href{http://www.arXiv.org/abs/1104.2863}{{\tt 1104.2863}}.

\bibitem{Armesto:2009fj}
N.~Armesto, L.~Cunqueiro, and C.~A. Salgado, ``{Q-PYTHIA: A Medium-modified
  implementation of final state radiation},'' {\em Eur.Phys.J.} {\bf C63}
  (2009) 679--690,
\href{http://www.arXiv.org/abs/0907.1014}{{\tt 0907.1014}}.

\bibitem{Gyulassy:1994ew}
M.~Gyulassy and X.-N. Wang, ``{HIJING 1.0: A Monte Carlo program for parton and
  particle production in high-energy hadronic and nuclear collisions},'' {\em
  Comput.Phys.Commun.} {\bf 83} (1994) 307,
\href{http://www.arXiv.org/abs/nucl-th/9502021}{{\tt nucl-th/9502021}}.

\bibitem{Ostapchenko:2004ss}
S.~Ostapchenko, ``{QGSJET-II: Towards reliable description of very high energy
  hadronic interactions},'' {\em Nucl.Phys.Proc.Suppl.} {\bf 151} (2006)
  143--146,
\href{http://www.arXiv.org/abs/hep-ph/0412332}{{\tt hep-ph/0412332}}.

\bibitem{Ahn:2009wx}
E.-J. Ahn, R.~Engel, T.~K. Gaisser, P.~Lipari, and T.~Stanev, ``{Cosmic ray
  interaction event generator SIBYLL 2.1},'' {\em Phys.Rev.} {\bf D80} (2009)
  094003,
\href{http://www.arXiv.org/abs/0906.4113}{{\tt 0906.4113}}.

\bibitem{Bopp:1998rc}
F.~W. Bopp, R.~Engel, and J.~Ranft, ``{Rapidity gaps and the PHOJET Monte
  Carlo},''
\href{http://www.arXiv.org/abs/hep-ph/9803437}{{\tt hep-ph/9803437}}.

\bibitem{Bopp:2005cr}
F.~W. Bopp, J.~Ranft, R.~Engel, and S.~Roesler, ``{Antiparticle to Particle
  Production Ratios in Hadron-Hadron and d-Au Collisions in the DPMJET-III
  Monte Carlo},'' {\em Phys.Rev.} {\bf C77} (2008) 014904,
\href{http://www.arXiv.org/abs/hep-ph/0505035}{{\tt hep-ph/0505035}}.

\bibitem{Werner:2010aa}
K.~Werner, I.~Karpenko, T.~Pierog, M.~Bleicher, and K.~Mikhailov,
  ``{Event-by-Event Simulation of the Three-Dimensional Hydrodynamic Evolution
  from Flux Tube Initial Conditions in Ultrarelativistic Heavy Ion
  Collisions},'' {\em Phys.Rev.} {\bf C82} (2010) 044904,
\href{http://www.arXiv.org/abs/1004.0805}{{\tt 1004.0805}}.

\bibitem{Werner:2007bf}
K.~Werner, ``{Core-corona separation in ultra-relativistic heavy ion
  collisions},'' {\em Phys.Rev.Lett.} {\bf 98} (2007) 152301,
\href{http://www.arXiv.org/abs/0704.1270}{{\tt 0704.1270}}.

\bibitem{Skands:2010ak}
P.~Z. Skands, ``{Tuning Monte Carlo Generators: The Perugia Tunes},'' {\em
  Phys.Rev.} {\bf D82} (2010) 074018,
\href{http://www.arXiv.org/abs/1005.3457}{{\tt 1005.3457}}.

\bibitem{Sjostrand:2004pf}
T.~Sj{\"o}strand and P.~Z. Skands, ``{Multiple interactions and the structure
  of beam remnants},'' {\em JHEP} {\bf 0403} (2004) 053,
\href{http://www.arXiv.org/abs/hep-ph/0402078}{{\tt hep-ph/0402078}}.

\bibitem{Sjostrand:2004ef}
T.~Sj{\"o}strand and P.~Z. Skands, ``{Transverse-momentum-ordered showers and
  interleaved multiple interactions},'' {\em Eur.Phys.J.} {\bf C39} (2005)
  129--154,
\href{http://www.arXiv.org/abs/hep-ph/0408302}{{\tt hep-ph/0408302}}.

\bibitem{Karneyeu:2013aha}
A.~Karneyeu, L.~Mijovic, S.~Prestel, and P.~Skands, ``{MCPLOTS: a particle
  physics resource based on volunteer computing},''
  \href{http://www.arXiv.org/abs/1306.3436}{{\tt 1306.3436}}.
{\url{http://mcplots.cern.ch}}.

\bibitem{Buckley:2010ar}
A.~Buckley, J.~Butterworth, L.~Lonnblad, H.~Hoeth, J.~Monk, {\em et al.},
  ``{Rivet user manual},''
\href{http://www.arXiv.org/abs/1003.0694}{{\tt 1003.0694}}.

\bibitem{d'Enterria:2011kw}
D.~d'Enterria, R.~Engel, T.~Pierog, S.~Ostapchenko, and K.~Werner,
  ``{Constraints from the first LHC data on hadronic event generators for
  ultra-high energy cosmic-ray physics},'' {\em Astropart.Phys.} {\bf 35}
  (2011) 98--113,
\href{http://www.arXiv.org/abs/1101.5596}{{\tt 1101.5596}}.

\bibitem{Martin:2009iq}
A.~Martin, W.~Stirling, R.~Thorne, and G.~Watt, ``{Parton distributions for the
  LHC},'' {\em Eur.Phys.J.} {\bf C63} (2009) 189--285,
\href{http://www.arXiv.org/abs/0901.0002}{{\tt 0901.0002}}.

\bibitem{Pumplin:2002vw}
J.~Pumplin, D.~Stump, J.~Huston, H.~Lai, P.~M. Nadolsky, {\em et al.}, ``{New
  generation of parton distributions with uncertainties from global QCD
  analysis},'' {\em JHEP} {\bf 0207} (2002) 012,
\href{http://www.arXiv.org/abs/hep-ph/0201195}{{\tt hep-ph/0201195}}.

\bibitem{Sherstnev:2008dm}
A.~Sherstnev and R.~Thorne, ``{Different PDF approximations useful for LO Monte
  Carlo generators},''
\href{http://www.arXiv.org/abs/0807.2132}{{\tt 0807.2132}}.

\bibitem{Buckley:2010jn}
A.~Buckley and M.~Whalley, ``{HepData reloaded: Reinventing the HEP data
  archive},'' {\em PoS} {\bf ACAT2010} (2010) 067,
\href{http://www.arXiv.org/abs/1006.0517}{{\tt 1006.0517}}.

\bibitem{Sandhoff:2005jh}
M.~Sandhoff and P.~Z. Skands, ``{Colour annealing - a toy model of colour
  reconnections},''.
FERMILAB-CONF-05-518-T, in hep-ph/0604120.

\bibitem{Skands:2007zg}
P.~Z. Skands and D.~Wicke, ``{Non-perturbative QCD effects and the top mass at
  the Tevatron},'' {\em Eur.Phys.J.} {\bf C52} (2007) 133--140,
\href{http://www.arXiv.org/abs/hep-ph/0703081}{{\tt hep-ph/0703081}}.

\bibitem{Schulz:2011qy}
H.~Schulz and P.~Skands, ``{Energy Scaling of Minimum-Bias Tunes},'' {\em
  Eur.Phys.J.} {\bf C71} (2011) 1644,
\href{http://www.arXiv.org/abs/1103.3649}{{\tt 1103.3649}}.

\bibitem{Donnachie:1992ny}
A.~Donnachie and P.~Landshoff, ``{Total cross-sections},'' {\em Phys.Lett.}
  {\bf B296} (1992) 227--232,
\href{http://www.arXiv.org/abs/hep-ph/9209205}{{\tt hep-ph/9209205}}.

\bibitem{Schuler:1993wr}
G.~A. Schuler and T.~Sj{\"o}strand, ``{Hadronic diffractive cross-sections and
  the rise of the total cross-section},'' {\em Phys.Rev.} {\bf D49} (1994)
2257--2267.

\bibitem{Abelev:2012sea}
{\bf ALICE} Collaboration, B.~Abelev {\em et al.}, ``{Measurement of inelastic,
  single- and double-diffraction cross sections in proton--proton collisions at
  the LHC with ALICE},'' {\em Eur.Phys.J.} {\bf C73} (2013) 2456,
\href{http://www.arXiv.org/abs/1208.4968}{{\tt 1208.4968}}.

\bibitem{Aamodt:2010pp}
{\bf ALICE} Collaboration, K.~Aamodt {\em et al.}, ``{Charged-particle
  multiplicity measurement in proton-proton collisions at $\sqrt{s}=7$ TeV with
  ALICE at LHC},'' {\em Eur.Phys.J.} {\bf C68} (2010) 345--354,
\href{http://www.arXiv.org/abs/1004.3514}{{\tt 1004.3514}}.

\bibitem{Aamodt:2010ft}
{\bf ALICE} Collaboration, K.~Aamodt {\em et al.}, ``{Charged-particle
  multiplicity measurement in proton-proton collisions at $\sqrt{s}=0.9$ and
  2.36 TeV with ALICE at LHC},'' {\em Eur.Phys.J.} {\bf C68} (2010) 89--108,
\href{http://www.arXiv.org/abs/1004.3034}{{\tt 1004.3034}}.

\bibitem{Corke:2010yf}
R.~Corke and T.~Sj{\"o}strand, ``{Interleaved Parton Showers and Tuning
  Prospects},'' {\em JHEP} {\bf 1103} (2011) 032,
\href{http://www.arXiv.org/abs/1011.1759}{{\tt 1011.1759}}.

\bibitem{Aad:2012mfa}
{\bf ATLAS} Collaboration, G.~Aad {\em et al.}, ``{Measurements of the
  pseudorapidity dependence of the total transverse energy in proton-proton
  collisions at $\sqrt{s}=7$ TeV with ATLAS},'' {\em JHEP} {\bf 1211} (2012)
  033,
\href{http://www.arXiv.org/abs/1208.6256}{{\tt 1208.6256}}.

\bibitem{Chatrchyan:2011wm}
{\bf CMS} Collaboration, S.~Chatrchyan {\em et al.}, ``{Measurement of energy
  flow at large pseudorapidities in $pp$ collisions at $\sqrt{s} = 0.9$ and 7
  TeV},'' {\em JHEP} {\bf 1111} (2011) 148,
\href{http://www.arXiv.org/abs/1110.0211}{{\tt 1110.0211}}.

\bibitem{Aspell:2012ux}
{\bf TOTEM} Collaboration, G.~Antchev {\em et al.}, ``{Measurement of the
  forward charged particle pseudorapidity density in $pp$ collisions at
  $\sqrt{s} = 7$ TeV with the TOTEM experiment},'' {\em Europhys.Lett.} {\bf
  98} (2012) 31002,
\href{http://www.arXiv.org/abs/1205.4105}{{\tt 1205.4105}}.

\bibitem{LombranaGonzalez:2012gd}
D.~Lombra{\~n}a~Gonzalez, F.~Grey, J.~Blomer, P.~Buncic, A.~Harutyunyan, {\em
  et al.}, ``{Virtual machines \& volunteer computing: Experience from
  LHC@Home: Test4Theory project},'' {\em PoS} {\bf ISGC2012} (2012)
036.

\end{thebibliography}\endgroup

\end{document}